# The LOFT Burst Alert System and its Burst On-board Trigger

Stéphane Schanne*[a], Diego Götz[a], Hervé Le Provost[b], Frédéric Château[b],
Enrico Bozzo[c], Søren Brandt[d],
on behalf of the LOFT collaboration

[a] CEA Saclay/IRFU/Service d'Astrophysique, 91191 Gif sur Yvette, France;
[b] CEA Saclay/IRFU/Service Electronique, Détecteurs et Informatique, 91191 Gif sur Yvette, France;
[c] ISDC, University of Geneva, Chemin d'Ecogia 16, 1290 Versoix, Switzerland;
[d] DTU Space, Elektrovej 327, 2800 Kongens Lyngby, Denmark.

## ABSTRACT

The ESA M3 candidate mission LOFT (Large Observatory For x-ray Timing) has been designed to study strong gravitational fields by observing compact objects, such as black-hole binaries or neutron-star systems and supermassive black-holes, based on the temporal analysis of photons collected by the primary instrument LAD (Large Area Detector), sensitive to X-rays from 2 to 50 keV, offering a very large effective area (>10 m$^2$), but a small field of view (ø<1°). Simultaneously the second instrument WFM (Wide Field Monitor), composed of 5 coded-mask camera pairs (2-50 keV), monitors a large part of the sky, in order to detect and localize eruptive sources, to be observed with the LAD after ground-commanded satellite repointing. With its large field of view (>π sr), the WFM actually detects all types of transient sources, including Gamma-Ray Bursts (GRBs), which are of primary interest for a world-wide observers community. However, observing the quickly decaying GRB afterglows with ground-based telescopes needs the rapid knowledge of their precise localization. The task of the Loft Burst Alert System (LBAS) is therefore to detect in near-real-time GRBs (about 120 detections expected per year) and other transient sources, and to deliver their localization in less than 30 seconds to the observers, via a VHF antenna network. Real-time full resolution data download to ground being impossible, the real-time data processing is performed onboard by the LBOT (LOFT Burst On-board Trigger system). In this article we present the LBAS and its components, the LBOT and the associated ground-segment.

## 1. SCIENTIFIC AND PROGRAMMATIC CONTEXT

The space mission LOFT (Large Observatory For x-ray Timing) [1] is an ESA candidate mission, dedicated to the study of black holes, neutrons stars and compact objects through their rapid variability in X-rays.

LOFT was one of the 4 missions (together with ECHO, STE-QUEST and MARCO-POLO-R) selected by ESA in February 2011 for an "Initial Assessment Study" (Phase A study) for the ESA M3 mission, to be launched in 2022 or 2024, depending on the schedule of the ESA L mission JUICE. The PLATO mission was also later added to the M3 competition. In February 2014 the final M3 mission selection took place and PLATO was selected. The LOFT proposal, judged of high quality by ESA, will be re-proposed for the M4 mission, with a target launch date in 2026. The LOFT mission's duration is nominally 3 years, with and extension of 2 additional years.

High time-resolution X-ray observations of compact objects provide direct access to strong-field gravity, black-hole masses and spins, and the equation of state of ultra-dense matter. A 10 m$^2$-class instrument in combination with good spectral resolution is required to exploit the relevant diagnostics and answer two fundamental questions of ESA's Cosmic Vision theme "Matter under extreme conditions", namely (1): does matter orbiting close to the black-hole event horizon follow the predictions of general relativity and (2): what is the equation of state of matter in neutron stars? Concerning black-holes, the LOFT scientific objectives are to measure their mass and spin, study their QPOs (Quasi-

Periodic Oscillations), test general relativity in the strong field domain and study the relativistic precession of matter in close orbit. Concerning neutron stars, LOFT will constrain their mass and radius to determine the equation of state of ultra-dense matter and study the properties of their crust. Furthermore, LOFT will operate as an observatory (during 50% of the time) and study all types of point-like sources relatively bright in X-rays, including X-ray bursters, High mass X-ray binaries, X-ray transients, Cataclysmic Variables, Magnetars and Gamma-ray bursts (GRBs), as well as nearby Active Galactic nuclei (AGNs).

Thanks to an innovative design and the development of large monolithic silicon drift detectors, the Large Area Detector (LAD) on-board LOFT achieves an effective area of ~10 m$^2$ (which is more than an order of magnitude larger than current space-borne X-ray detectors and the future ATHENA mission) in the 2-30 keV range (up to 80 keV in extended mode) with 10 μs timing, yet still fitting on a conventional platform and a small/medium-class launcher. With this large area and a spectral resolution better than 260 eV over its entire energy band, LOFT will be capable to revolutionize the study of collapsed objects, yielding unprecedented information on strongly curved space-time and matter under extreme conditions of pressure and magnetic field strength.

Besides the LAD, LOFT will be carrying the Wide Field Monitor (WFM, Figure 1) [2], whose goal is to monitor in the 2-50 keV energy band the entire sky region (90°×180°+90°×90°) accessible by the LAD, and to inform the scientists in case an interesting transient source (e.g. a black hole binary, a millisecond pulsar, etc.), is changing state and becoming bright and interesting to be studied in detail by the LAD, after ground repointing of the satellite.

The LAD and the WFM are based on the same detector technology, namely Silicon Drift Detectors (SDDs). While the LAD has no imaging capabilities and a small field of view of 1°, the WFM can map a large part of the high-energy sky thanks to the coded mask technique. The WFM allows localization of persistent and transient sources to about 1 arcmin (for a 100 mCrab source during 1 ks exposure). At the same time, the sources localized by the WFM will be studied with unprecedented energy resolution (< 500 eV at 6 keV) and sensitivity (5 mCrab for 50 ks exposure).

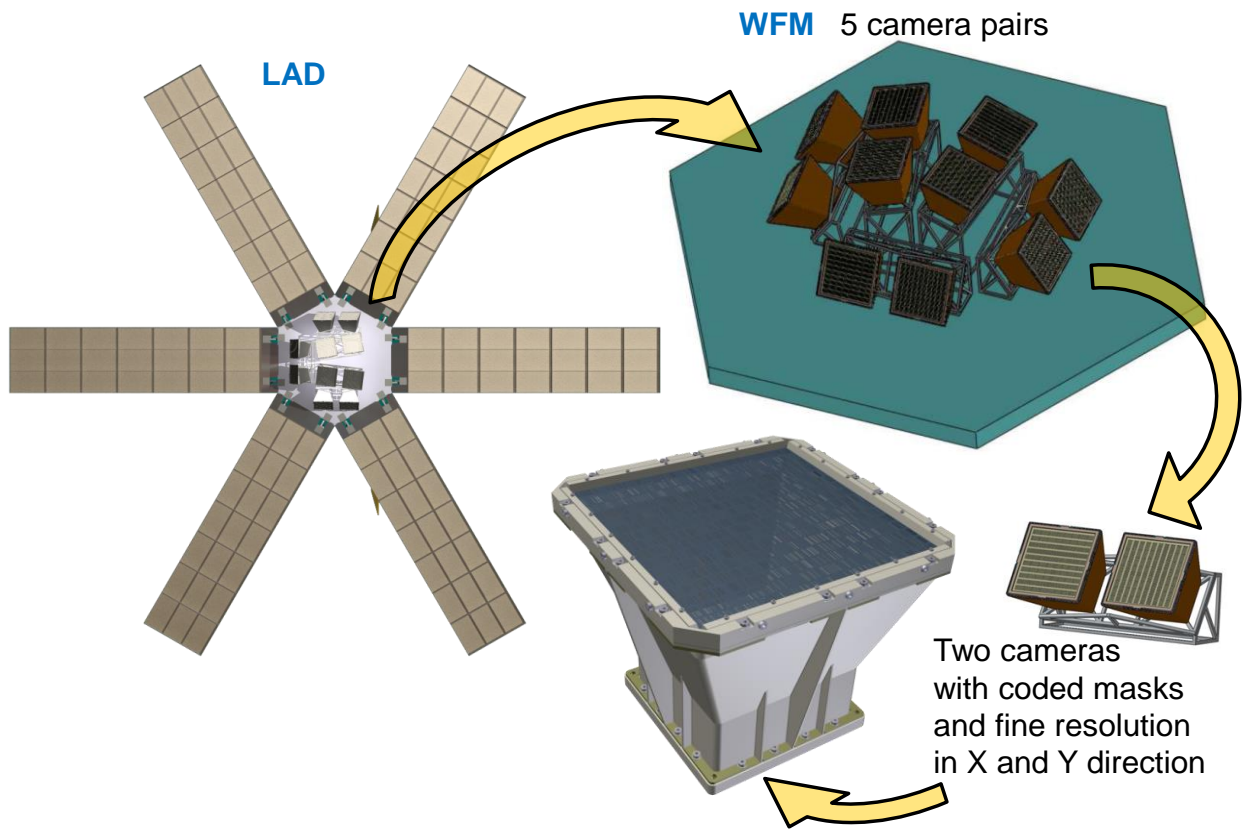

Figure 1: Scheme of the WFM instrument on-board the LOFT spacecraft.

| Parameter | Value | Comment |
|---|---|---|
| Field of view | 4.5 sr | 180° x 90° + 90° x 90° anti-solar direction, reduced by Earth occultation in LEO. |
| Effective area max | 80 cm$^2$ | on axis, for a camera pair |
| Energy | 2 – 30 keV | up to 50 keV extended |
| Energy resolution | ~500 eV | SDD optimized for imaging |
| Timing resolution | 10 µs | |
| Spatial resolution | 5 arcmin | Localization ~ 1 arcmin |
| Sensitivity | 1 Crab in 1 s | 5 mCrab in 50 ks |

Figure 2: WFM key parameters (Brand et al [3])

## 2. SCIENCE WITH LOFT EXTENDED TO A REAL-TIME ASTRONOMY MISSION

The WFM on-board LOFT, thanks to its key parameters (Figure 2), in particular its wide field of view, good localization accuracy and sensitivity, is a prime instrument for the study of fast astronomical transients in near real-time. This aspect has been added to the original LOFT proposal during the assessment phase, to increases the mission's scientific return and to involve a larger European and worldwide scientific community, at relatively low cost without compromise on the main scientific objectives of LOFT.

In order to investigate the physics of fast astronomical transients, among which primarily Gamma-Ray Bursts (GRBs), past satellite missions have shown that a multi-wavelength follow-up of those events is mandatory. The Italian-Dutch BeppoSAX mission provided for the first time GRB positions within a delay of a few hours and allowed to associate these elusive flashes of gamma-ray emission with the death of massive stars, through the detection of their afterglow in X-rays, optical and radio bands. In addition, ground based observations allowed for the first time to measure the distance (redshift) of a subset of the GRB sample allowing to establish their cosmological nature. The era of real-time astrophysics really began with the successors of BeppoSAX, which are the missions HETE-II, INTEGRAL and Swift which collected a database comprising more than a hundred GRBs with measured redshift.

| Events per Year | Transient source type |
|---|---|
| ~ 120 | GRBs detected by WFM |
| ~ 25 | Luminous GRBs allowing precise timing and spectroscopy |
| ~ 2<br>< 1 | High redshift GRBs (z>6)<br>- detected by WFM<br>- with measured redshift |
| ~ 40 | XRFs (X-ray flashes) |
| ~ thousands<br>~ hundreds<br>~ some | type-I X-ray bursts,<br>magnetar flares,<br>SN shock breakout,<br>tidal disruption events, … |

Figure 3: Expected WFM per year detection rates (from [4], following [5]).

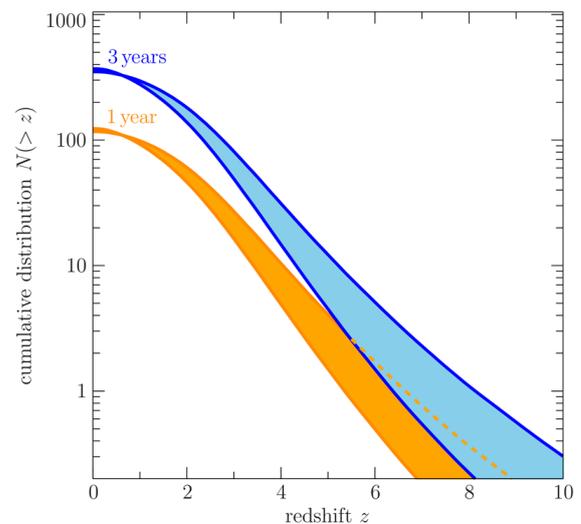

Figure 4: Expected GRB detections by WFM as a function of z (from [4], following [5] and [6]).

With the LOFT Burst Alert System (LBAS), dedicated to the real-time detection of rapid transient sources on-board the WFM, the LOFT scientific objectives now include this field of astrophysics. The number of GRBs expected to be detected by the WFM can be evaluated, based on the detection rates of present and past GRB experiments flying on satellites in LEO (BATSE, Swift) and by taking into account the different field of view and sensitivity of WFM with respect to those instruments. Using the logN-logS and logN-logP of GRBs (BATSE, Swift) and XRFs (BeppoSAX-WFC, HETE-2) and taking into account the Earth occultation effect, the estimated GRB detection rates by the WFM are presented in Figure 3 (following Amatti et al [5]). For the GRB detection rate as a function of redshift (Figure 4), different possible luminosity functions of GRBs were calibrated against Swift GRBs and convolved with the WFM field of view and sensitivity (method by Salvaterra et al [6]).

As we can see, thanks to its very large field of view, we expect to detect a large number of GRBs with the WFM. Their good localization accuracy (of about ~1 arcmin) is well adapted to follow-up observations. However, the intermediate step of performing localization refinements of those WFM GRBs by ground-based robotic telescopes down to the arcsec level, needed by large spectroscopic telescopes because of their small field of view, cannot be avoided by the WFM. Indeed unlike other missions (e.g. Swift and SVOM), LOFT does not offer autonomous slews to repoint onboard follow-up telescopes onto the source to perform these localization refinements. Therefore the number of actually measured redshifts for the WFM GRBs is expected to remain small compared to dedicated GRB missions.

On the other hand, with its very low energy-threshold of 2 keV, the WFM will do excellent GRB science and offer major improvements in the field. The WFM will detect a large number of X-ray rich GRBs. Indeed GRBs observed by BeppoSAX and Hete-II (which offered similar low-energy thresholds) where relatively short at high energies, while having extended tails with many photons at low energies, which is a hint that GRBs have much to reveal at low energies. The WFM will also study intensively X-Ray Flashes (XRFs), a subclass of faint and very soft GRBs, which could actually form the majority of them.

Furthermore it has been recognized that, in order to advance on our understanding of the GRB phenomenon itself, after the years of extended afterglow studies with the Swift-XRT, we should study more in detail the prompt emission in the low-energy band. This is what the WFM will do. Overlaid with the generally accepted model of GRB photons being produced by synchrotron emission, some models predict thermal and comptonized components, detectable at low energy in the WFM band, a prime goal for this mission.

Possible detections of transient X-ray absorption features in GRB spectra by WFM (BeppoSAX had already observed some of them in a few GRB cases), could give new insights on the circum-stellar burst medium and maybe on the progenitor, and possibly allow direct redshift measurements using the line shift of the absorption feature (which can be crosschecked with the redshift measured by afterglow observations with optical means).

Furthermore with its low energy-threshold, WFM will have access to the domain of the very high redshift bursts, and bring advancements in cosmological studies with GRBs. This will allow the determination of the rate of star formation and the rate of massive star explosions in the young/distant universe, as well as the study of the cosmic evolution of the interstellar medium. To be able to perform these studies, redshift measurements by WFM GRB afterglow follow-up observations with other means than LOFT are however needed.

To allow those GRB follow-up observations, the LOFT Burst Alert System (LBAS) is built. Il will deliver quick GRB position alerts as a fundamental service to the scientific community. Indeed, at the 2026 horizon, LOFT may well be the only mission capable of delivering GRB alerts (after missions like Swift and SVOM being already terminated). Particularly on this horizon the scientific community will be very much interested in GRB alerts, since many missions will perform GRB follow-up observations (on ground in the visible/IR domain with ELT and LSST, in radio with SKA and at ultra-high energies with CTA; in space in the IR domain with JWST and in X-rays with LOFT-LAD) or simultaneous detections of the GRB prompt emission (with wide-angle cameras in the visible, or with the new powerful detectors of non-electromagnetic messengers such as neutrinos and gravitational waves).

## 3. THE LOFT BURST ALERT SYSTEM (LBAS)

The LOFT Burst Alert System (LBAS) aims at detecting and localizing transient events in real-time and informing the scientific community as quickly as possible (within ~ 30 s) about the location of the event on the sky (with ~ 1 arcmin accuracy), to allow rapid follow-up observations in a world-wide joint effort. Rapid-response robotic telescopes, receiving the LBAS alerts, will automatically repoint onto the event localized by LBAS. Within their field of view of typically tens of arcmin, they will search for the optical counterpart and refine its localization to arcsec-level, which is of the size of the field of view of large telescopes, which in turn will perform spectroscopic studies of the afterglow, to fully characterize the event and determine its redshift and hence distance.

Such an observing strategy has been followed by the three missions HETE-II, INTEGRAL and Swift. Different solutions have been used to provide near-real-time GRB alerts to the community : HETE-II detected the transients on-board and sent only positional information to ground over a VHF permanent link through a network of ground-receiver stations located in the Earth equatorial zones under its projected track; INTEGRAL sends its complete data set over a permanent real-time data link to ground, permitting near real-time data analysis on ground; Swift detects transients on-board and sends their positional information to ground over the TDRSS satellite relay-system.

The implementation scheme chosen for the LBAS is similar to the HETE-II one, with an on-board detection and alert generation system coupled to an on-board VHF emitter and a ground-receiver VHF network. Indeed, since LOFT is a low-Earth-orbit mission, the INTEGRAL solution is excluded because the full data-set cannot be transmitted in real-time to ground; and the Swift solution is not envisaged, since it relies on a strictly non-European system.

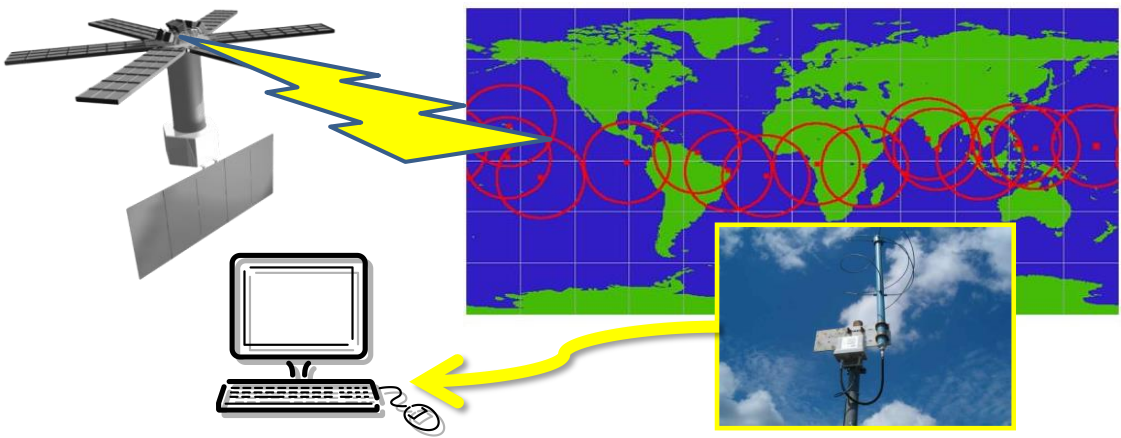

Figure 5: LBAS scheme - including LBOT, VHF emitter, VHF receivers and Alert Centre.

The LBAS is made of the following four key parts (Figure 5):

- The LBOT (LOFT Burst Onboard Trigger system), an on-board transient source detection and localization system. Its function is to process the WFM camera data in real-time in order to detect and localize the appearance of a new transient source on the sky. Whenever a new source has been found, the LBOT's two main actions are (i) to generate a sequence of real-time alert messages describing the source (including the time and position of the event), and (ii) to secure in the on-board memory the full resolution data produced by the WFM during the timespan covering minutes before to minutes after the burst. Those data are stored in photon by photon mode for later download (giving access to full energy and timing accuracy, while the standard WFM data are energy-binned histograms accumulated over typically tens of seconds, losing the fine time resolution).

- An on-board VHF emitter system, which serializes the LBOT alert message sequences and broadcasts them over a VHF antenna to ground. This system implements a permanent low-bandwidth (~600 bit/s) data downlink.

- A network of VHF ground-receiver stations spread under the satellite track on Earth. With the nearly equatorial orbit (inclination < 5°) and low altitude (~600 km) of the satellite, about 15 VHF receiver stations placed under the satellite track in the Earth equatorial zone are needed. They are composed of a VHF antenna connected to a HF receiver, demodulator and digitizing computer connected to the internet. Those stations are designed as simple and reliable systems, with remote control and reset capabilities. They are often located in remote areas with minimal radio-interferences, and are operated by personnel linked to the scientific community.

- The LAC (LOFT Alert Centre) [7], which is the on-ground alert distribution center. It automatically collects and orders the alert message packets received from the various VHF stations, performs quality checks, builds the alert notices, and posts them to the scientific community over the internet in the usual form of GCN (Gamma-Ray Burst Coordinate Network) notices or VOE (Virtual Observatory Events). A human "burst advocate" is finally responsible of manually confirming each automatic alert notices. The maximal delay between alert generation onboard and automatic alert message dissemination to the community is as short as 20 s in 66% of the cases. Longer delays are expected in case of a non-working VHF receiver station due to atmospheric scintillation, radio-interference or internet outage. The VHF onboard emitter thus repeats each message several times in order to get it once the satellite has moved in reach of the next VHF receiver station.

A similar triggering and alert dissemination system is currently under development for the SVOM mission (the Space based Variable Objects Monitor, see e.g. Schanne et al [8]), a French-Chinese mission dedicated to GRB studies. Indeed, onboard SVOM, a system with functions similar to those of the LBOT, called the UTS system (Unité de Traitement Scientifique, see Schanne et al [9] and Le Provost et al [10]), is developed at Irfu/CEA Sacaly. It analyses in real-time the data stream of SVOM's large field-of-view gamma-ray detector ECLAIRs in order to detect and localize transient events, to (i) initiate a satellite slew maneuver in order to place the transient source into the field of view of two narrow-field instruments on-board SVOM (a visible and an X-ray telescope) for rapid afterglow follow-up observations by the SVOM mission itself, and to (ii) produce VHF alert message sequences similar to those foreseen on LOFT, to be distributed to the community.

The SVOM mission is scheduled to be operational in orbit before LOFT, therefore the LBOT can fully benefit from the developments for the SVOM/UTS, as well as the SVOM/VHF emitter, the equatorial part of the SVOM/VHF receiver-station system, and parts of the SVOM/on-ground alert distribution center, which can be reused for the LOFT mission.

## 4. THE LOFT BURST ON-BOARD TRIGGER (LBOT)

The LBOT (LOFT Burst On-board Trigger system), developed at Irfu/CEA Saclay, is the first stage of the LBAS, whose function is to process the WFM camera data in real-time in order to detect and localize the appearance of a new transient source on the sky, among which primarily GRBs. The detection of a new transient source produces a VHF-alert message sequence containing the source localization; on a large time-span around the event the photon by photon data of the triggered WFM cameras are secured for later downlink over high-bandwidth telemetry for on-ground offline analysis.

The LBOT system is part of the Data Handling Unit (DHU) [11] of the LOFT WFM. The DHU collects the data of all 5 pairs of coded-mask cameras of the WFM (see Figure 1). Each of those cameras employs a 1D-coded mask placed over a detection plane made of Silicon Drift Detectors, with coarse resolution along the drift direction and fine positioning along the anodes direction which is parallel to the coded direction of the mask. Both cameras of each pair are coaxial and rotated by 90° with respect to each other around the optical axis, such that 2D source positioning is achieved by combination of both camera images.

GRBs show large variations in duration, as well as large variability in their light curves. The LBOT triggers therefore on time scales ranging from 10 ms to 20 min. Three energy bands are used by the LBOT trigger in order to be sensitive to all kinds of GRBs: a low energy band (typically 2 to 8 keV) for X-ray rich GRBs, X-ray Flashes and high redshifted GRBs; a higher energy band (typically 8 to 32 keV) for standard GRBs; and the full energy range for faint events. Since most GRBs appear off axis, the LBOT triggers not only on the full detector plane, but uses 9 zones (the full plane, 4 halves and 4 quadrants). Since full sky image reconstruction cannot be systematically performed for all cameras on the shortest time scales due to lack of computation power, two types of trigger algorithms, similar to those foreseen for SVOM (Schanne et al [12]), are implemented:

- an "Image Trigger", which performs systematic sky imaging on long duration time-scales (e.g. > 20 s).
- a "Count-Rate Increase Trigger", which in a first step carefully selects a time-scales on which a GRB candidate is detected, followed by a second step in which this time-scales is imaged.

To avoid a large number of triggers on false or uninteresting events, on-board source filtering is used. Celestial sources rising above the Earth horizon are identified knowing the spacecraft orientation. Activities from known sources are identified and ignored using a known sources catalogue on-board. Strong known sources are filtered-out upfront the trigger algorithms using mask-tagging to remove the pixels exposed by such sources.

The "Image Trigger" algorithm is devoted to the search of long duration transient events, above typically 20 seconds. Its principle is to build, on a cyclic time-basis, the detector plane images of all 10 cameras, to model and subtract the spatial shape of the backgrounds induced by partial obscuration by the Earth, to deconvolve those images using the mask patterns to produce sky images, which in turn are summed to build sky images of all long duration time-scales considered (typically 20 s to 20 min). In those sky images, known-source positions and portions obscured by the Earth are filtered out, excesses in terms of signal-to-noise ratio are searched for, and coincidences in both images of a camera pair are matched (Figure 6), resulting in a new-source detection. The peaks in both images of the camera pair are fit, giving two fine positions in orthogonal directions, which permit to derive a 2D localization at arcmin-level on the sky, used to produce the VHF alert message.

The "Count-Rate Increase Trigger" algorithm is devoted to the search of short duration transient events, below typically 20 seconds. It is based on a two steps algorithm. In a first step are searched and stored into a buffer all significant count-rate increases over the estimated background in both cameras of each WFM camera pair, on each energy band, detector zone and time-interval from 10 ms to 20 s. In a second step the best excess present in the buffer is searched, the detector plane images of both cameras in a pair are built, deconvolved to produce the two corresponding sky images, in which matching excesses are searched, vetoed against the known source catalogue, and fit to derive the 2D localization, used to produce the VHF alert message.

It should be noticed that the use of a camera pair instead of a single camera, permits to trigger on much fainter sources, thanks to the capability to perform coincidence detections between both cameras.

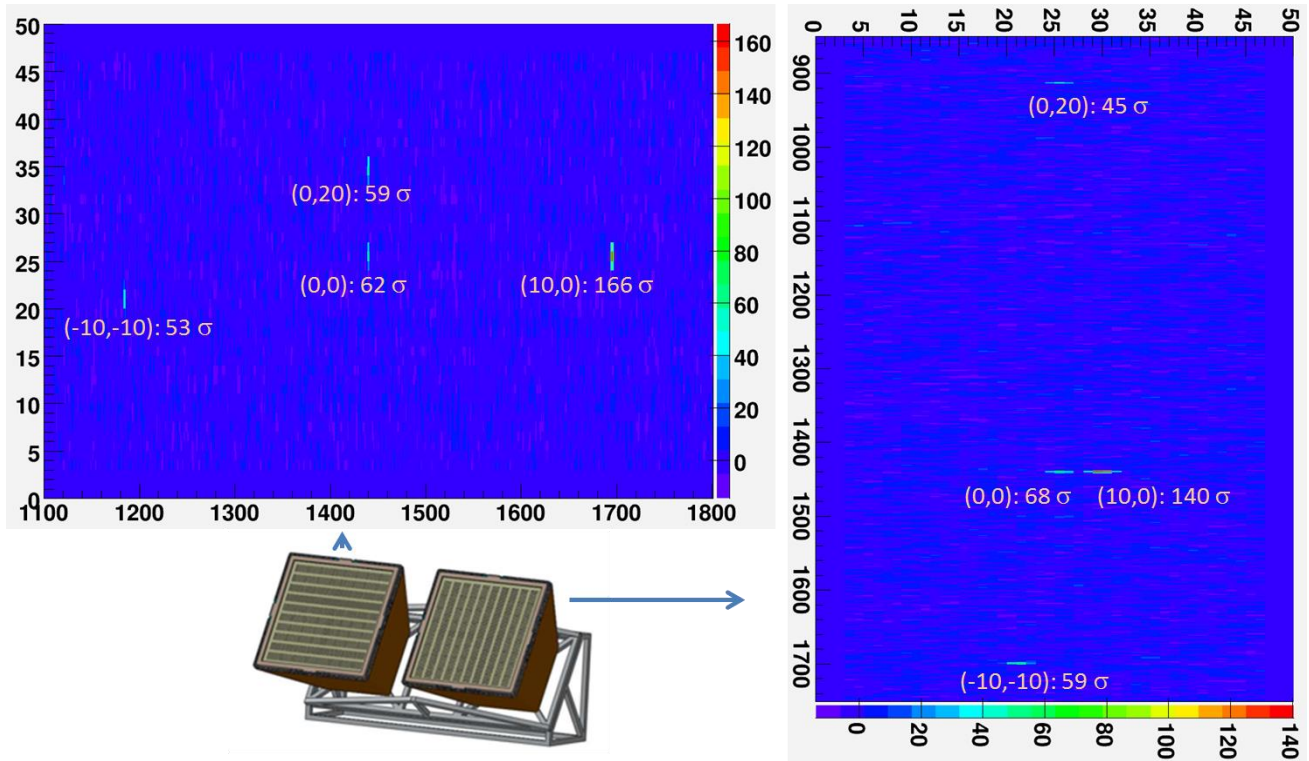

Figure 6: Central portions of sky images, in units of signal to noise ratio (number of sigma), obtained by a WFM camera-pair imaging-simulation software, developed in order to test the imaging capability of such a device. In this example 4 sources, located at the coordinates (0°,0°), (10°,0°), (0°,20°) and (-10°,-10°) with respect to the common optical axis of the camera pair, and with fluxes of 15, 5, 5 and 5 photons/cm$^2$/s, are projected during a 100 s exposure, through the masks onto the detection planes of each camera pair, together with the expected cosmic extragalactic background. The masks (of size ~26 cm) have 1040×16 physical pixels, from which 25% are transparent to the X-rays, and arranged in a random pattern. The detector planes (of size ~15 cm, placed ~20 cm below the masks) have been binned into 1024×18 pixels (of size 0.14×8.08 mm derived from the resolution in the anode and drift direction). For the deconvolution process the masks are rebinned onto this detector pixel grid, resulting in sizes of 1856×32 pixels and sky images of 2880×50 pixels. The detector plane image deconvolution with the rebinned masks produces two sky images from which, for clarity reasons, only the central parts are shown in the figure. All 4 sources are detected in both cameras, with fine localization on the horizontal axis for the first camera and on the vertical axis for the second camera, from which a fine 2D localization is derived by matching the peaks in both images. Furthermore, as tests with low intensity sources have shown, taking peaks in both images in coincidence permits to use a much lower triggering threshold (in terms of signal to noise ratio) than with one camera alone.

The LBOT is foreseen to be implemented as a system embedded inside the DHU hardware, mainly a very powerful FPGA at the core of the DHU. The LBOT is based on two main parts:

- a Firmware part (in VHDL code), whose function is to manage the LBOT interface with the rest of the DHU. In particular this firmware performs in real-time the pre-processing and binning of the data produced by the 10 WFM Cameras, and the storage into memory of those computation results. It also manages the emission of the alert messages produced, as well as the production of the system house-keeping data for trigger algorithm control. The Firmware part also implements a powerful CPU as an IP module residing inside the FPGA.

- a Software part (written in C++), which runs on the CPU implemented by the Firmware in the processing FPGA. This software runs the triggering algorithms, which use the data pre-processed by the Firmware part, stored in memory, to seek for the appearance of a new unknown transient source on the sky, performs its localization, with the subsequent generation of the VHF alert messages.

To develop the LBOT algorithms, studies are ongoing to model the data expected from the WFM as an input stream to the LBOT. Those studies benefit from developments currently carried out for SVOM. They are based on GRB catalogues from previous missions and synthetic GRB tables, as well as background models taking into account the mission's orbit and the Earth occultations. The LBOT algorithms are developed in such a way, they can be implemented both in a standalone computer with large CPU power, allowing the triggering efficiency to be evaluated on large GRB catalogues, and in the target hardware, allowing their benchmarking and the assessments of the needed CPU power.